\documentclass[
    ,final            
    ,numberedheadings 
  ]
  {aipproc}

\usepackage{natbib}

\layoutstyle{6x9}

\newcommand\aj{AJ}%
\newcommand\apj{ApJ}%
\newcommand\apjl{ApJL}%
\newcommand\aap{A\&A}%
\newcommand\pasp{PASP}%
\newcommand\mnras{MNRAS}%

\begin{document}

\title{The Supernova Type Ia Rate Evolution with SNLS}


\author{James~D.~Neill}{
  address={University of Victoria, PO Box 3055, Victoria, BC V8W 3P6, Canada}
}

\author{M.~Sullivan}{
  address={University of Toronto, 50 St. George Street, Toronto, ON M5S 3H4, Canada}
}

\author{D.~Balam}{
  address={University of Victoria, PO Box 3055, Victoria, BC V8W 3P6, Canada}
}

\author{C.~J.~Pritchet}{
  address={University of Victoria, PO Box 3055, Victoria, BC V8W 3P6, Canada}
}

\author{D.~A.~Howell}{
  address={University of Toronto, 50 St. George Street, Toronto, ON M5S 3H4, Canada}
}
 
\author{K.~Perrett}{
  address={University of Toronto, 50 St. George Street, Toronto, ON M5S 3H4, Canada}
}
 
\author{P.~Astier}{
  address={LPNHE, CNRS-IN2P3 and University of Paris VI \& VII, 75005 Paris, France}
}
 
\author{E.~Aubourg}{
  address={APC, 11 Pl. M. Berthelot, 75231 Paris Cedex 5, France},
  altaddress={DSM/DAPNIA, CEA/Saclay, 91191 Gif-sur-Yvette Cedex, France}
}

\author{S.~Basa}{
  address={LAM CNRS, BP8, Traverse du Siphon, 13376 Marseille Cedex 12, France}
}
 
\author{R.~G.~Carlberg}{
  address={University of Toronto, 50 St. George Street, Toronto, ON M5S 3H4, Canada}
}

\author{A.~Conley}{
  address={University of Toronto, 50 St. George Street, Toronto, ON M5S 3H4, Canada}
}
 
\author{S.~Fabbro}{
  address={CENTRA IST, Avenida Rovisco Pais, 1049 Lisbon, Portugal}
}

\author{D.~Fouchez}{
  address={CPPM, CNRS-IN2P3 and Univ. Aix Marseille II, Case 907, 13288 Marseille Cedex 9, France}
}
 
\author{J.~Guy}{
  address={LPNHE, CNRS-IN2P3 and University of Paris VI \& VII, 75005 Paris, France}
}

\author{I.~Hook}{
  address={University of Oxford Astrophysics, Denys Wilkinson Building, Keble Road, Oxford OX1 3RH, UK}
}
 
\author{R.~Pain}{
  address={LPNHE, CNRS-IN2P3 and University of Paris VI \& VII, 75005 Paris, France}
}

\author{N.~Palanque-Delabrouille}{
  address={DSM/DAPNIA, CEA/Saclay, 91191 Gif-sur-Yvette Cedex, France}
}
 
\author{N.~Regnault}{
  address={LPNHE, CNRS-IN2P3 and University of Paris VI \& VII, 75005 Paris, France}
}

\author{J.~Rich}{
  address={DSM/DAPNIA, CEA/Saclay, 91191 Gif-sur-Yvette Cedex, France}
}
 
\author{R.~Taillet}{
  address={LPNHE, CNRS-IN2P3 and University of Paris VI \& VII, 75005 Paris, France},
  altaddress={Universit\'{e} de Savoie, 73000 Chamb\'ery, France}
}

\author{G.~Aldering}{
  address={LBNL, 1 Cyclotron Rd, Berkeley, CA 94720, USA}
}
 
\author{P.~Antilogus}{
  address={LPNHE, CNRS-IN2P3 and University of Paris VI \& VII, 75005 Paris, France}
}

\author{V.~Arsenijevic}{
  address={CENTRA IST, Avenida Rovisco Pais, 1049 Lisbon, Portugal}
}
 
\author{C.~Balland}{
  address={APC, 11 Pl. M. Berthelot, 75231 Paris Cedex 5, France}
}

\author{S.~Baumont}{
  address={LPNHE, CNRS-IN2P3 and University of Paris VI \& VII, 75005 Paris, France}
}
 
\author{J.~Bronder}{
  address={University of Oxford Astrophysics, Denys Wilkinson Building, Keble Road, Oxford OX1 3RH, UK}
}

\author{R.~S.~Ellis}{
  address={California Institute of Technology, E. California Blvd., Pasadena, CA 91125, USA}
}
 
\author{M.~Filiol}{
  address={LAM CNRS, BP8, Traverse du Siphon, 13376 Marseille Cedex 12, France}
}

\author{A.~C.~Gon\c{c}alves}{
  address={LUTH,UMR 8102, CNRS and Observatoire de Paris, F-92195 Meudon, France},
  altaddress={CAAUL, Observat\'orio Astron\'omico de Lisboa, Tapada da Ajuda, 1349-018 Lisbon, Portugal}
}

\author{D.~Hardin}{
  address={LPNHE, CNRS-IN2P3 and University of Paris VI \& VII, 75005 Paris, France}
}

\author{M.~Kowalski}{
  address={LBNL, 1 Cyclotron Rd, Berkeley, CA 94720, USA}
}
 
\author{C.~Lidman}{
  address={ESO, Alonzo de Cordova 3107, Vitacura, Casilla 19001, Santiago 19, Chile}
}

\author{V.~Lusset}{
  address={DSM/DAPNIA, CEA/Saclay, 91191 Gif-sur-Yvette Cedex, France}
}
 
\author{M.~Mouchet}{
  address={APC, 11 Pl. M. Berthelot, 75231 Paris Cedex 5, France}
}

\author{A.~Mourao}{
  address={CENTRA IST, Avenida Rovisco Pais, 1049 Lisbon, Portugal}
}
 
\author{S.~Perlmutter}{
  address={LBNL, 1 Cyclotron Rd, Berkeley, CA 94720, USA}
}

\author{P.~Ripoche}{
  address={CPPM, CNRS-IN2P3 and Univ. Aix Marseille II, Case 907, 13288 Marseille Cedex 9, France}
}
 
\author{D.~Schlegel}{
  address={LBNL, 1 Cyclotron Rd, Berkeley, CA 94720, USA}
}

\author{C.~Tao}{
  address={CPPM, CNRS-IN2P3 and Univ. Aix Marseille II, Case 907, 13288 Marseille Cedex 9, France}
}

\begin{abstract} We present a progress report on a project to derive the
evolution of the volumetric supernova Type Ia rate from the Supernova
Legacy Survey.  Our preliminary estimate of the rate evolution divides
the sample from \citet{Neill06AJ} into two redshift bins: $0.2 < z <
0.4$, and $0.4 < z < 0.6$.  We extend this by adding a bin from the
sample analyzed in \citet{Sullivan06ApJ} in the range $0.6 < z < 0.75$
from the same time period.  We compare the derived trend with
previously published rates and a supernova Type Ia production model
having two components: one component associated closely with star
formation and an additional component associated with host galaxy
mass.  Our observed trend is consistent with this model, which
predicts a rising SN~Ia rate out to at least $z=2$.

\end{abstract}

\maketitle


\section{Introduction}

The importance of Type Ia supernovae (SNe~Ia) as cosmic distance tracers
\cite{Riess98AJ,Perlmutter99ApJ} motivates the efforts to understand their
progenitors.  We need not only explain the physics that makes their
explosions so useful, but we must also recognize and control the
systematic effects that may result from the properties of their progenitors
and their evolution.

One method used to explore various progenitor models is to compare the
evolution of star formation in the universe with the evolution of the
observed SN~Ia rate \cite{Cappellaro99A&A, Hardin00A&A, Pain02ApJ,
Madgwick03ApJ, Tonry03ApJ, Blanc04A&A, Dahlen04ApJ, Barris06ApJ}.  The
challenge of this method is that it requires minimizing systematic effects
that can produce spurious trends as a function of redshift.  The redshift
regions most important for constraining the progenitors, the highest and
lowest redshifts, are also the most prone to systematic effects.   At
low redshifts cosmic variance and the difficulty of sampling large
volumes and thus produce good statistics are severe challenges.  At high
redshifts, the faintness of the objects produce low signal-to-noise
detections and less certain typing as well as higher host contamination
since projected host offsets are smaller.

The Supernova Legacy Survey (SNLS) offers the opportunity to measure the
SN~Ia rate trend right in the sweet spot where the volume sampled per
square degree is large and yet the objects are bright enough for high
quality typing.  This circumstance motivated the recently published study
of \citet{Neill06AJ} which produced an anchor point for rate evolution
studies at $z\sim0.5$.  Our next step is to extend this effort by binning
the original sample and adding a higher redshift bin to trace the rate
evolution from $z=0.2$ to $z=0.75$.  The homogeneous sample and high
quality typing provided by SNLS can thus achieve a reduction in the
systematics in this redshift range and provide a foundation for rate
evolution studies at higher and lower redshifts.

Even though rigorous constraints of SN~Ia progenitors await accurate
low and high redshift rate measurements, we can still address a few
questions that have arisen from other studies of SN~Ia rates.  We
would like to see if we can reproduce the observed trends from
previously published studies, in particular those that show a large
jump in the SN~Ia rate just beyond $z=0.5$ \cite{Barris06ApJ}.  Of
particular importance to future SN~Ia surveys, such as JDEM and LSST,
is the question of the rate evolution beyond $z=1$, which appears to
decline dramatically \cite{Dahlen04ApJ}.  This seems to contradict the
observation in the local universe that the SN~Ia rate is higher in
star forming galaxies \cite{Cappellaro99A&A, Oemler79AJ,
  vandenBergh90PASP, Mannucci05A&A} which implies that the SN~Ia rate
should increase at least out to $z\sim2$.  This correlation of SN~Ia
rate with host star formation rate has now been observed in the
intermediate redshift universe using SNLS \cite{Sullivan06ApJ} which
implies it is not a local effect.  For this study we ask: is the
observed trend in the SN~Ia volumetric rate consistent with this
correlation of rate with host properties?

\section{The Supernova Legacy Survey}

The SNLS was instigated with the goal of providing observational
constraints on the pressure/density ratio of the universe, $w$.  Its
first-year results \cite{Astier06A&A}, in combination with other
cosmological probes, provide one of the best observational constraints
on $w$ \cite{Spergel06}.  It is designed as a rolling-search survey of
four one-square-degree fields evenly spaced in RA.  Each field is
imaged with the Megacam imager on the Canada-France-Hawaii Telescope
five times a month during its roughly six-month observing season in
four wavelength bands: {\it g'r'i'z'}.  A deep {\it u*} image is also
taken of each field to aid in measuring host galaxy properties.  As
each epoch is acquired it is scanned for variable objects which are
recorded in a database\footnote{see
  \url{http://legacy.astro.utoronto.ca}, and
  \url{http://makiki.cfht.hawaii.edu:872/sne/}}.  SN-like variables
are assessed for their SN~Ia type likelihood \cite{Sullivan06AJ} and
all useful SN~Ia candidates are promoted for followup spectroscopy
with the VLT or the Gemini or Keck Telescopes.  The availability of 8
and 10 meter class telescopes for spectroscopic followup provides a
well defined completeness and a large spectroscopically confirmed
sample of objects.  The database of all variable objects provides a
means to estimate spectroscopic completeness \cite{Neill06AJ}.

\section{Rate Evolution with SNLS}

\citet{Neill06AJ} used a Monte Carlo method to estimate the SNLS
survey efficiency in the redshift range $0.2 < z < 0.6$.  Updated host
extinction models \cite{Riello05MNRAS} improved the estimation of the
systematic errors.  The study of SN~Ia rates as a function of host
properties by \citet{Sullivan06ApJ} used the same sample and methods,
but added an additional sample in the range $0.6<z<0.75$.

To estimate the rate evolution using the SNLS, we split the sample
from \cite{Neill06AJ} and \cite{Sullivan06ApJ} into three bins:
$0.2<z<0.4$, $0.4<z<0.6$, and $0.6<z<0.75$.  Separate Monte Carlo runs
were performed for each sub-sample including experiments to determine
the associated systematic errors in each redshift bin.  Table 1 lists
the spectroscopically confirmed sample for each field in each redshift
bin.  The preliminary results reported here will be improved upon by
adding more recent SNe~Ia from the survey to increase the sample size
in each bin.

We applied the same methods used in \cite{Neill06AJ} to derive the
volumetric SN~Ia rate in each bin, and associated systematic and
statistical errors. We now compare the results to the two-component
model of SN~Ia production \cite{Sullivan06ApJ, Mannucci05A&A,
  Scannapieco05ApJ}.

\begin{table}
\begin{tabular}{lrrr}
\hline
 \tablehead{1}{r}{b}{Field} & \tablehead{1}{r}{b}{$0.2<z<0.4$} &
 \tablehead{1}{r}{b}{$0.4<z<0.6$} & \tablehead{1}{r}{b}{$0.6<z<0.75$} \\
\hline
D1  &  3 & 13 &  9 \\
D2  &  5 & 10 & 12 \\
D3  &  6 & 10 & 12 \\
D4  &  3 &  8 &  9 \\
\hline
ALL & 17 & 41 & 42 \\
\hline
\end{tabular}
\caption{Spectroscopically confirmed SN~Ia samples}
\end{table}

\section{The Two-Component Model}

\begin{figure}[h]
  \includegraphics[height=0.60\textheight,angle=90.]{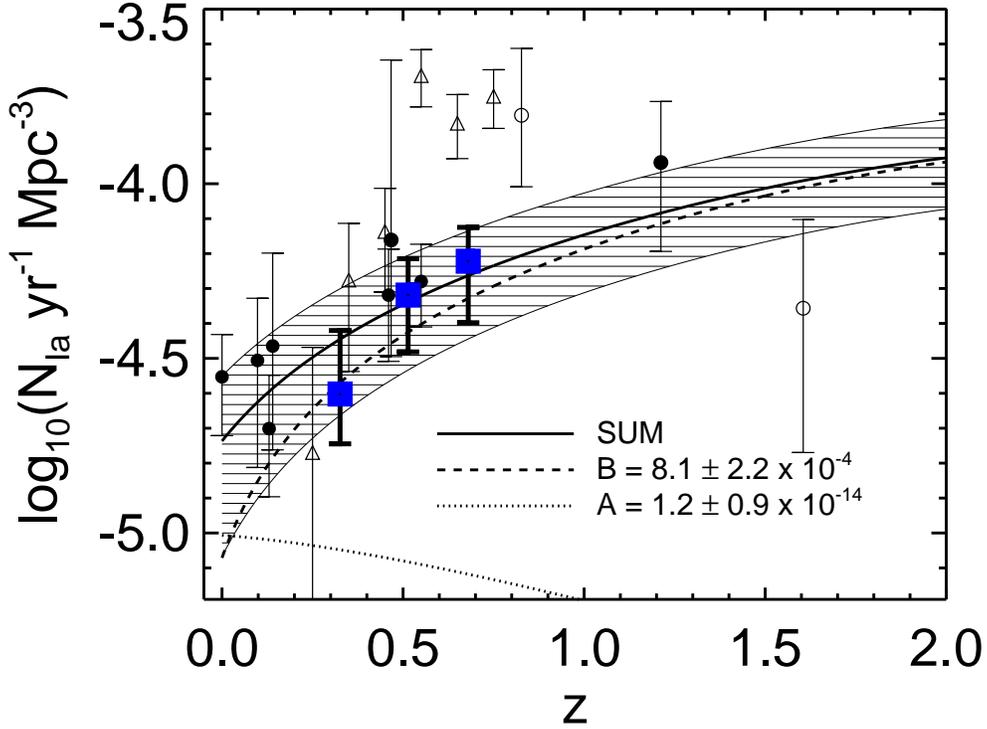}
  \caption{Observed SNLS SN~Ia volumetric rate evolution. The filled
    squares are the SNLS rates in each of the three redshift bins.
    The filled circles are previously published SN Ia rates derived
    from samples primarily confirmed by
    spectroscopy from the following references (in redshift order):
    \cite{Cappellaro99A&A,Madgwick03ApJ,Blanc04A&A,Hardin00A&A,
      Tonry03ApJ,Dahlen04ApJ,Pain02ApJ,Dahlen04ApJ}.  The open circles
    are the SN~Ia rates from \cite{Dahlen04ApJ} with samples having
    only 50\% spectroscopic confirmation.  The open triangles are the
    rates from \cite{Barris06ApJ}, whose samples are confirmed almost
    entirely with photometric methods.  Error bars represent the
    systematic and statistical errors added in quadrature.}
\end{figure}

This model proposes a component associated with the host galaxy mass,
to explain the non-zero SN~Ia rate observed in early-type galaxies,
and a component associated with the host galaxy star formation rate,
to explain the correlation of SN~Ia rate with host star formation rate
\cite{Sullivan06ApJ, Mannucci05A&A}.  The volumetric SN~Ia rate
evolution, $r_V(t)$, is thus expressed as follows: \begin{equation}
  r_V(t) = A M_*(t) + B \dot{M}_*(t), \label{eq_ab}\end{equation} with
$A$ in terms of SNe~Ia per year per unit host stellar mass and $B$ in
terms of SNe~Ia per year per unit host star formation.  The $B$
component produces SNe~Ia on shorter ($\sim 1$Gyr) time-scales, while
the $A$ component is responsible for the longer time-scale required to
produce SNe~Ia in galaxies dominated by older stellar populations.

Figure 1 shows a fit (solid line) of the SNLS rate evolution (filled
squares) and various rates from the literature (solid circles) to the
two-component model based on the star formation history (SFH) as
parameterized by \citet{Hopkins06}.  The resulting coefficients are
annotated on the plot and the resulting evolution of each component is
indicated by the dashed line ($B$ component) and the dotted line ($A$
component).  The statistical uncertainty in the combined components
is indicated by the hashed region.

We note that this model appears to be incompatible with two features
in the observed rates.  This fit is statistically inconsistent with
the large jump in the rates seen at $z=0.5$ by \citet{Barris06ApJ}.
It also predicts an increase in the SN~Ia rate out to $z>2$, while the
rates from \citet{Dahlen04ApJ} show a steady decline after a redshift
of $z\sim 1$.  The highest redshift observation ($z=1.6$) is only
marginally inconsistent with the model.

The mass used in fitting the $A$ and $B$ components was derived by
integrating the SFH from \citet{Hopkins06}, and thus includes the mass
of dead stars.  When fitting host properties, the mass derived does
not include dead stars.  Our fitted values for $A$ and $B$ are
consistent with published values \citep[\S6.1.1]{Neill06AJ}, but are
marginally offset from host property values as expected from the
differences in host masses.  Our final results will include a mass
correction for dead stars and thus produce $A$ and $B$ values more
consistent with those derived from host properties.

\section{Conclusions}

By restricting our measurements of the SN~Ia volumetric rate evolution
to samples whose members are primarily identified with spectroscopy,
we find a trend that appears compatible with the observed properties
of SN~Ia hosts.  The most stringent test of this model awaits further
measurements of the rate beyond $z=1$, but we must be careful.  If the
two-component model is correct, the higher redshift SNe~Ia are more
closely associated with star formation.  This close association could
introduce systematics from the dust and other features of
high-redshift star formation that compound the difficulty of deriving
accurate high-redshift SN~Ia rates.  This caution applies to the use
of these high-redshift SNe~Ia for cosmology as well.


\begin{theacknowledgments}
The authors wish to recognize and acknowledge the very significant cultural
role and reverence that the summit of Mauna Kea has always had within the
indigenous Hawaiian community.  We are grateful for our opportunity to
conduct observations on this mountain.  We acknowledge invaluable
assistance from the CFHT Queued Service Observations team, led by P. Martin
(CFHT).  Our research would not be possible without the assistance of the
support staff at CFHT, especially J.-C. Cuillandre.  The real-time
pipelines for supernovae detection run on computers integrated in the CFHT
computing system, and are very efficiently installed, maintained and
monitored by K.  Withington (CFHT).  We also heavily rely on the real-time
Elixir pipeline which is operated and monitored by J.-C. Cuillandre, E.
Magnier and K.  Withington.  We are grateful to L. Simard (CADC) for
setting up the image delivery system and his kind and efficient responses
to our suggestions for improvements.  The Canadian collaboration members
acknowledge support from NSERC and CIAR; French collaboration members from
CNRS/IN2P3, CNRS/INSU, PNC and CEA.  This work was supported in part by the
Director, Office of Science, Office of High Energy and Nuclear Physics, of
the US Department of Energy.  The France-Berkeley Fund provided additional
collaboration support.  We are indebted to A. Hopkins and J. Beacom for
providing us with a draft of their work on SFH prior to its publication.
The views expressed in this article are those of the author and do not
reflect the official policy or position of the United States Air Force,
Department of Defense, or the U.S. Government.
\end{theacknowledgments}



\bibliographystyle{aipproc}   



\end{document}